\documentclass{pasa}%

\title[The S-PASS total intensity source catalogue]{A southern-sky total intensity source catalogue at 2.3\,GHz from S-band Polarisation All-Sky Survey data}
\author[B. W. Meyers et al.]
{B. W. Meyers$^\mathrm{A,B,C,}$\thanks{email: bradley.meyers@postgrad.curtin.edu.au}, 
N. Hurley-Walker$^\mathrm{A}$, 
P. J. Hancock$^\mathrm{A,B}$, 
T. M. O. Franzen$^\mathrm{A}$, 
E. Carretti$^\mathrm{C,D}$, 
L. Staveley-Smith$^\mathrm{E,B}$, 
B. M. Gaensler$^\mathrm{F,B}$,
M. Haverkorn$^\mathrm{G,H}$,
S. Poppi$^\mathrm{D}$
\\
\affil{$\mathrm{^A}$International Centre for Radio Astronomy Research (ICRAR), Curtin University, Bentley, WA 6102, Australia}
\affil{$\mathrm{^B}$ARC Centre of Excellence for All-Sky Astrophysics (CAASTRO)}
\affil{$\mathrm{^C}$CSIRO Astronomy and Space Science, P.O. Box 76, Epping, New South Wales 1710, Australia}
\affil{$\mathrm{^D}$INAF / Osservatorio Astronomico di Cagliari, Via della Scienza 5, I-09047 Selargius, Italy}
\affil{$\mathrm{^E}$International Centre for Radio Astronomy Research (ICRAR), The University of Western Australia, Crawley, WA 6009, Australia}
\affil{$ \mathrm{^F} $Dunlap Institute for Astronomy and Astrophysics, 50 St. George St, University of Toronto, ON M5S 3H4, Canada}
\affil{$ \mathrm{^G} $Department of Astrophysics / IMAPP, Radboud University Nijmegen, PO Box 9010, 6500 GL Nijmegen, the Netherlands}
\affil{$ \mathrm{^H} $Leiden Observatory, Leiden University, PO Box 9513, 2300 RA Leiden, the Netherlands}
}%
\jid{PASA}
\doi{10.1017/pas.\the\year.xxx}
\jyear{\the\year}

\usepackage[authoryear]{natbib}
\bibpunct{(}{)}{;}{a}{}{,}
\usepackage{aas_macros}
\usepackage{hyperref} 
\hypersetup{colorlinks,citecolor=blue,linkcolor=blue,urlcolor=blue}
\usepackage{lscape}

\newcommand{\spass}{\mbox{S-PASS}}
\newcommand{\nsrc}{23,389}
\newcommand{\spasssky}{16,600}
\newcommand{\rms}{12.9}

\begin{document}%
\begin{abstract}
The S-band Polarisation All-Sky Survey (\spass) has observed the entire southern sky using the 64-metre Parkes radio telescope at $ 2.3\mathrm{\,GHz} $ with an effective bandwidth of $ 184\mathrm{\,MHz} $.
The surveyed sky area covers all declinations $ \delta\leq 0^\circ $. 
To analyse compact sources the survey data have been re-processed to produce a set of 107 Stokes \textit{I} maps with $ 10.75\mathrm{\,arcmin} $ resolution and the large scale emission contribution filtered out. 
In this paper we use these Stokes \textit{I} images to create a total intensity southern-sky extragalactic source catalogue at $ 2.3\mathrm{\,GHz} $. 
The source catalogue contains \nsrc\ sources and covers a sky area of $ \spasssky\mathrm{\,deg^2} $, excluding the Galactic plane for latitudes $ |b|<10^\circ $. 
Approximately $ 8\% $ of catalogued sources are resolved.
\spass\ source positions are typically accurate to within $ 35\mathrm{\,arcsec} $. 
At a flux density of $ 225\mathrm{\,mJy} $ the \spass\ source catalogue is more than $ 95\% $ complete, and $ \sim 94\% $ of \spass\ sources brighter than $ 500\mathrm{\,mJy\,beam^{-1}} $ have a counterpart at lower frequencies.
\end{abstract}
\maketitle
\section{INTRODUCTION}
\label{sec:intro}
Radio source catalogues that cover wide areas of sky are important tools for exploring the properties and evolution of a large range of source populations.
Combining multiple source catalogues allows the determination of source spectral index information and statistical studies with large samples of different radio galaxy populations, such as active galactic nuclei (AGN) and star-burst galaxies.
Measuring how the relative fractions of different populations change and, ultimately, how the differential source counts evolve with frequency (see \citealp{2010A&ARvdeZotti} for a review) provides essential insight into the co-evolution of galaxies and their central super-massive black holes through cosmic time.

The S-band Polarisation All-Sky Survey (\spass) has mapped the southern sky for declinations $ \delta\leq 0^\circ $ in total intensity and polarisation with the 64-metre Parkes radio telescope at a frequency of $ 2300\,\mathrm{MHz} $. 

\spass\ is a project to map the diffuse emission of the entire southern sky at $ 2.3\mathrm{\,GHz} $. 
The main survey goals are to investigate the polarised synchrotron emission, Galactic and extragalactic magnetism, and Cosmic Microwave Background polarised foregrounds.
A more detailed description of the \spass\ survey strategy and science goals is given by \citet{2010ASPCCarretti} and in the upcoming survey description paper (Carretti et al., in prep.).  

One \spass\ data product is a collection of all-southern-sky total intensity maps, containing more than $ 10^4 $ extragalactic radio sources. 
Most of the radio sources in the \spass\ images will be distant radio galaxies.  

\renewcommand{\arraystretch}{1.1}
\begin{table*}
\caption{Radio source catalogues used in the comparison and verification of the \spass\ source catalogue.}
\begin{center}
\begin{tabular}{@{}lcccccc@{}}
\hline\hline
Survey catalogue & Frequency & Resolution & Flux density limit & Epoch & Overlap area & $ N_\mathrm{sources}^\dagger $  \\
 & [GHz] & [arcmin] & $5\sigma_\mathrm{rms}$ [$ \mathrm{mJy\,beam^{-1}} $] &  & [$\times 1000\mathrm{\,deg}^2$] & \\ 
\hline
S-PASS & 2.3 & 10.75 & 65 & 2007--2010 & 16.6 & \nsrc \\ 
SUMSS & 0.843 & $\sim 0.75$ & 8--18 & 1997--2003 & 7 & 209,186 \\ 
NVSS & 1.4 & 0.75 & 11 & 1993--1996 & 11 & 567,556 \\  
PMN$^a$ & 4.85 & 4.2 & 20--45 & 1990 & 8.6 & 17,297 \\ 
PKSCAT90 & 2.7 & $\sim 6$ & $\sim 50$ & 1990 & 16.3 & 5,884 \\
ATCA calibrators$^b$ & 2.1 & $\sim 0.1 $ & -- & 2016 & -- & 363 \\
\hline\hline
\end{tabular}
\end{center}
\tabnote{$^\dagger$This is the number of sources in the overlap region only and excludes sources in the comparison catalogues that fall within the $ |b|<10^\circ $ cut imposed on the \spass\ source catalogue. If declination cuts are imposed during verification, they are explicitly stated in the text.} 
\tabnote{$^a$The full PMN catalogue is comprised of four sub-catalogues. For this paper, we use only the ``Southern'' and ``Zenith'' sub-catalogues.}
\tabnote{$^b$Selected sources above $ 500\mathrm{\,mJy} $ within the nominal \spass\ declination range.}
\label{tab:survey_comp}
\end{table*}

In this paper, we present the construction and verification of the \spass\ Stokes \textit{I} source catalogue. 
We compare the \spass\ source catalogue to several other radio source catalogues to assess its quality (see Table~\ref{tab:survey_comp}): the Sydney University Molonglo Sky Survey (SUMSS; \citealp{2003MNRASMauch}); the NRAO VLA Sky Survey (NVSS; \citealp{1998AJCondon}); the Parkes-MIT-NRAO survey (PMN; \citealp{1993AJGriffith} and corresponding paper series); the Australia Telescope Compact Array (ATCA) calibrator list\footnote{http://www.narrabri.atnf.csiro.au/calibrators/} and; the Parkes Radio Source Catalogue (PKSCAT90; \citealp{1979AuJPABolton,1990PKSWright}).

The paper is structured as follows. 
In Section \ref{sec:obs_and_reduct} the observation strategy and image processing is outlined. 
In Section \ref{sec:construction} we describe the procedures used to construct the source catalogue. 
Section \ref{sec:analysis} contains the analysis and verification of the catalogue and in Section \ref{sec:cat_format} we outline the catalogue format. 
Finally, we review our conclusions in Section \ref{sec:summary}. 
Throughout, spectral indices, $ \alpha $, are defined using the convention $ S_\nu\propto\nu^\alpha $.

\section{DATA COLLECTION \& REDUCTION}
\label{sec:obs_and_reduct}
\subsection{Observations}
\label{sec:observations}
Observations were carried out over the period October 2007 to January 2010 using the Parkes S-band receiver. 
The S-band receiver is a package with: a system temperature $ T_\mathrm{sys}=20\mathrm{\,K} $, a beam Full Width at Half Maximum (FWHM) of $ 8.9\mathrm{\,arcmin} $, and a circular polarisation front-end ideal for linear polarisation observations with single-dish telescopes. 

Observing was carried out in long azimuth scans taken at the elevation of the south celestial pole as viewed from Parkes covering the entire declination range ($ \delta\leq 0^\circ$) in each scan.
Specifically, a scan length in azimuth of $ 115^\circ $ and a scan rate of $ 15\,\mathrm{deg\, min^{-1}} $ is required to realise this. 
Earth rotation was used to span the whole RA range. 

As described in \citet{2010ASPCCarretti}, each night a zig-zag in the sky is realised (see Figure 8 of \citealp{2010ASPCCarretti}). 
Combining the different zig-zags taken on different nights, all of the RA range can be observed with the appropriate sampling.
The azimuthal scans are observed either eastward at sky-rise, or westward at sky-set.
That way, two full sets of scans are realised with different directions in the sky, that, combined, provide an effective basket weaving.  

The Parkes observatory staff performed pointing calibrations at the beginning of each session, delivering the telescope with pointing offsets better than $ 10\mathrm{\,arcsec} $ in both RA and Dec (more than sufficient for $ 9\mathrm{\,arcmin} $ beam-width observations).
Scans to check pointing calibrations were performed at each session by the observing team to check that no residual offset along the scan direction was present.
More details can be found in \citet{2010ASPCCarretti}, while a full description will be included in the forthcoming \spass\ survey description paper (Carretti et al., in prep.).

Data were collected with the Digital Filter Bank mark 3 (DFB3) using a configuration with 256 MHz bandwidth and 512 frequency channels ($ 0.5\,\mathrm{MHz} $ channel width). This configuration also provides full Stokes information (autocorrelation products for the two circular polarisations RR* and LL*, and their complex cross-product RL*). 

The primary flux density calibrator was PKS B1934-638, using the model from \citet{1994Reynolds}, with PKS B0407-658 as the secondary calibrator. 
The resulting absolute flux calibration is accurate to within $ 5\text{--}10\% $.

\subsection{Data reduction}
\label{sec:reduction}
A software pipeline developed by the \spass\ team was employed to reduce and calibrate the data given the complex observation strategy and science goals.
Output data were binned into $ 8\mathrm{\,MHz} $ channels for calibration and radio frequency interference (RFI) flagging purposes. 
The calibrator flux density model was used to calibrate each individual channel, giving a flat calibrated bandpass (see \citealp{2013MNRASCarretti}) without need for further corrections.
After RFI flagging, the useful band covered the ranges \mbox{2176--$ 2216\mathrm{\,MHz} $} and \mbox{2256--$ 2400\mathrm{\,MHz} $}. 
All useful $ 8\mathrm{\,MHz} $ bands were binned together in one channel for an effective central frequency of $ 2307\mathrm{\,MHz} $ and $ 184\mathrm{\,MHz} $ bandwidth.

The maps are arranged in rings of declination such that the entire \spass\ observed sky is captured.
Map centres are $ -7.5^\circ $, $ -22.5^\circ $, $ -37.5^\circ $, $ -52.5^\circ $, $ -67.5^\circ $ and $ -82.5^\circ $ -- with 24, 24, 21, 18, 13 and 7 maps respectively in each declination range. 
Each map is a grid of $ 3\times 3\mathrm{\,arcmin^2} $ in zenithal equidistant (ARC) projection.
For analysis focused on compact sources, \spass\ scans are spatially high-pass filtered to remove the large scale spatial structure \citep{2016ApJLamee}. 
A median filter with a $ 45\mathrm{\,arcmin} $ window was used to achieve this.
A window size of $ 5\times $ the intrinsic resolution ($ 9\mathrm{\,arcmin} $) was chosen in order to give the best trade-off between ineffectively removing large scale structure and not affecting the source flux estimates.
The filtered scans were then spatially convolved with a $ 6\mathrm{\,arcmin} $ Gaussian window.
All data points within the Gaussian window were binned and weighted based on the window function value at that pixel coordinate.
This generated the final set of 107 $ 15 \times 15\mathrm{\,deg^2} $ maps, with an effective beam width of $\theta_\mathrm{fwhm} = 10.75\mathrm{\,arcmin} $.

The mean RMS noise in the Stokes \textit{I} maps is $ \sigma_\mathrm{rms}\approx \rms\mathrm{\,mJy\,beam^{-1}} $. 
Since the thermal noise is an order of magnitude lower ($ \sigma_\mathrm{th}\approx 1\mathrm{\,mJy\,beam^{-1}} $, \citealp{2013NatCarretti}), the sensitivity is limited by the confusion noise which we estimate to be $\sigma_\mathrm{c}=\sqrt{\sigma_\mathrm{rms}^2-\sigma_\mathrm{th}^2}\approx \rms\mathrm{\,mJy\,beam^{-1}} $. 
This is consistent with the estimate from a scaled approximation of equation (14) in \citet{1974ApJCondon}, 
\begin{equation}
\sigma_\mathrm{c}\approx 0.2\left(\frac{\nu}{\mathrm{GHz}}\right)^{-0.7}\left(\frac{\theta_\mathrm{fwhm}}{\mathrm{arcmin}}\right)^2\approx 12.9\mathrm{\,mJy\,beam^{-1}}
\end{equation} 
which is appropriate for synthesised beams larger than $\theta_\mathrm{fwhm}= 0.17\mathrm{\,arcmin} $.

\section{CATALOGUE CONSTRUCTION}
\label{sec:construction}
The final \spass\ source catalogue was constructed by combining the source catalogue for each of the 107 total intensity maps. 
Here we detail the catalogue creation process for one tile and then how the individual tile catalogues were combined to create the final \spass\ source catalogue.

\begin{figure}
\centering
\includegraphics[width=0.5\textwidth]{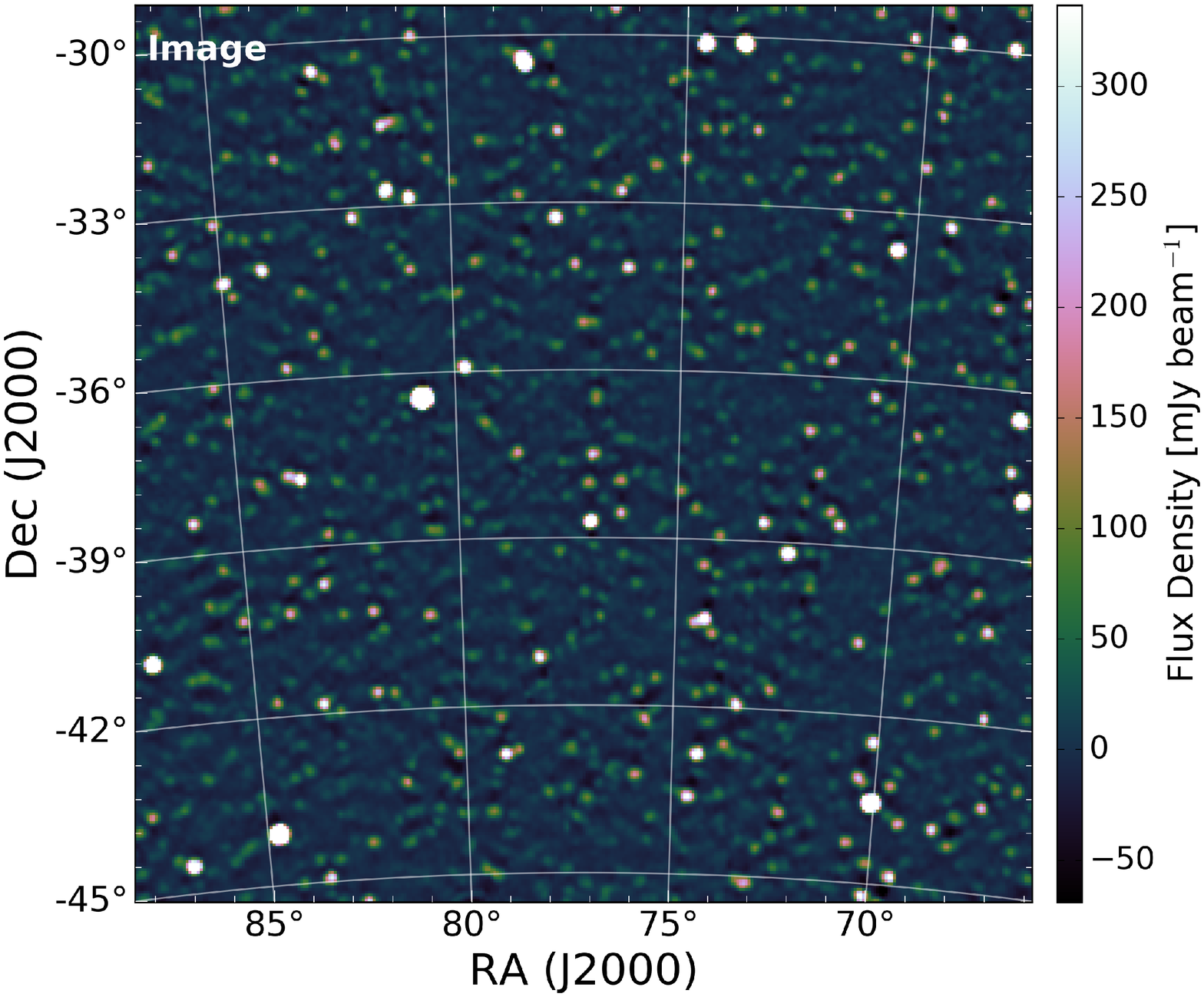}
\includegraphics[width=0.51\textwidth]{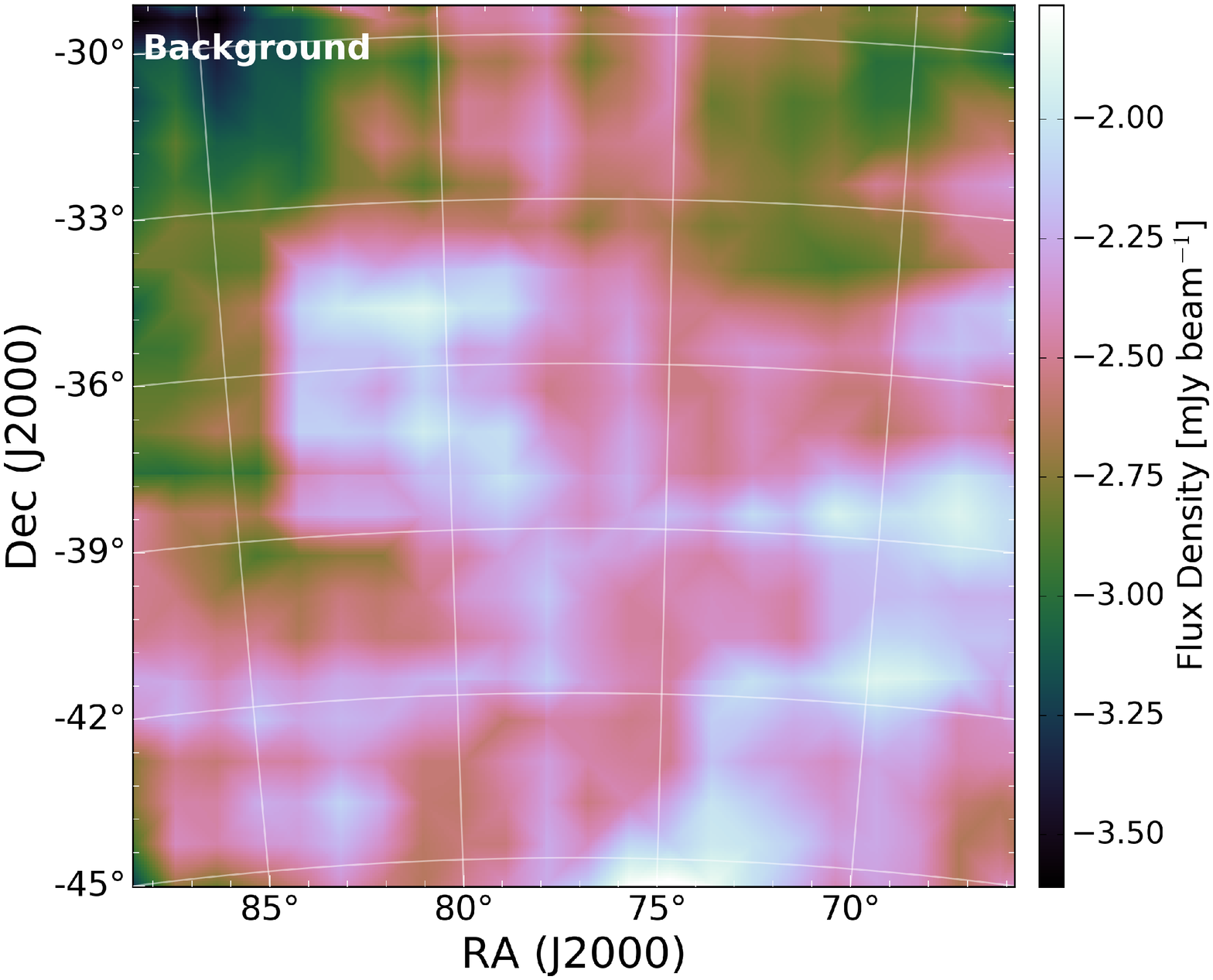}
\includegraphics[width=0.49\textwidth]{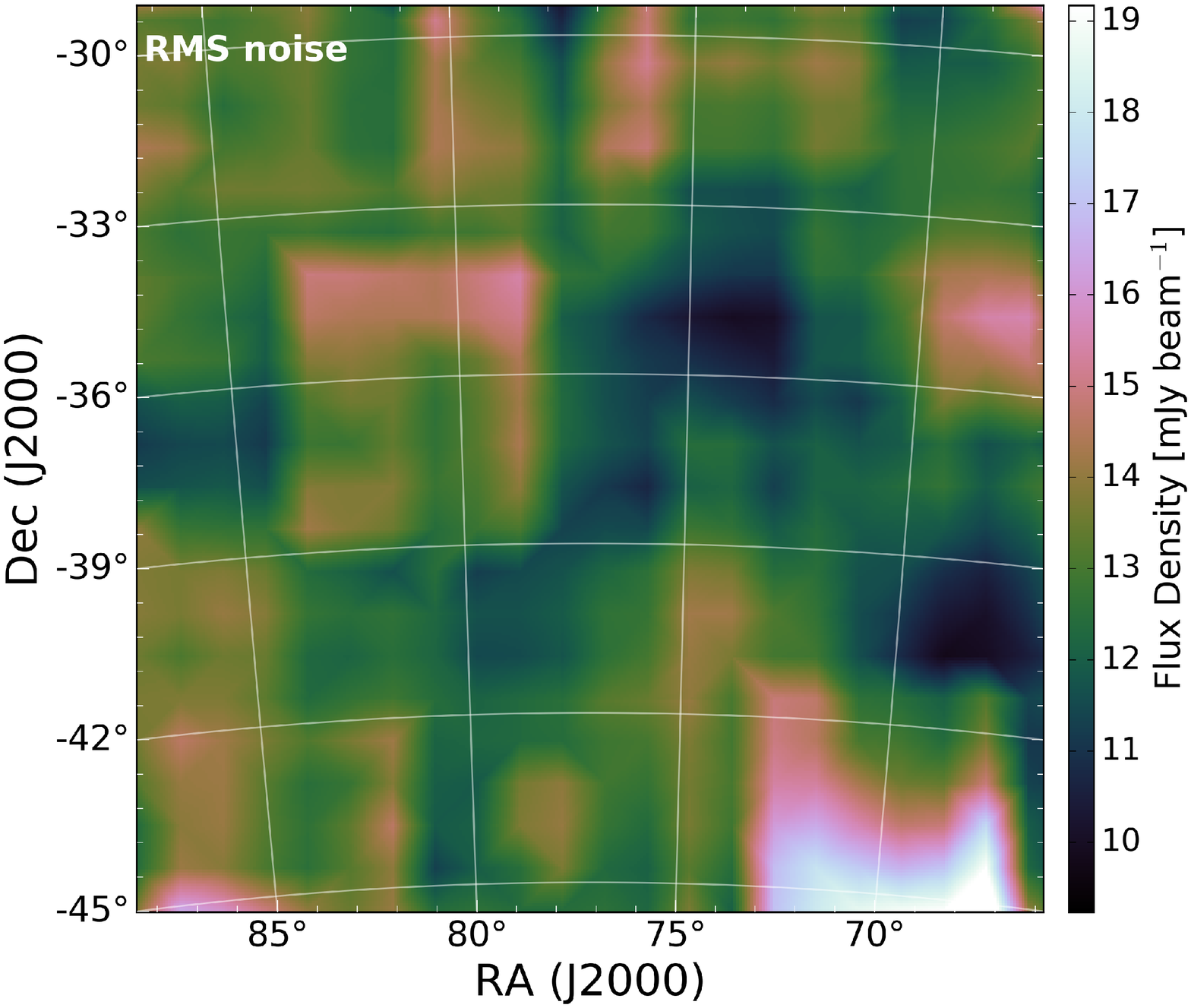}
\caption{\textit{Top}: A typical \spass\ image, centred on J2000 coordinates $(\alpha,\delta)=$ (05:08:42, -37:31:30). \textit{Middle}: The background estimation for the image produced by \textsc{bane}. Values are negative due to the median filtering applied (see Section~\ref{sec:reduction}). \textit{Bottom}: The RMS noise map produced by \textsc{bane}.}
\label{fig:img_bkg_rms}
\end{figure}

\subsection{Source finding for one tile}
\label{sec:srcfind}
We used the source finding algorithm \textsc{aegean}\footnote{v2.0b-81-g6b1142c-(2016-09-08), see {http://ascl.net/1212.009}} \citep{2012MNRASHancock} and its associated tool set to create a source catalogue from the raw images. 
\textsc{aegean} fits one or more elliptical Gaussians to each source and produces a set of characterising source parameters.
A signal-to-noise cut of $ 5\sigma_\mathrm{src} $, where $ \sigma_\mathrm{src} $ is the local RMS for each source as calculated during the noise estimation step, was imposed on the tile catalogues.

\subsubsection{Background and noise estimation}
The Background And Noise Estimation tool (\textsc{bane}; part of the \textsc{aegean} tool set) was used to create background and RMS noise maps, evaluated on angular scales of $ \sim 3^\circ$, for individual images. 
See Figure~\ref{fig:img_bkg_rms} for an example of an \spass\ tile image and the corresponding background and RMS noise maps.
Background values for the individual maps are expected to be close to zero, if slightly negative, due to the median filtering applied to the images. 
Typical background values measured by \textsc{bane} are $ \approx -2.3\mathrm{\,mJy\,beam^{-1}} $, but range from $ -3.7\mathrm{\,mJy\,beam^{-1}} $ to $ 0.5\mathrm{\,mJy\,beam^{-1}} $. 
The combination of the median filtering, described in Section~\ref{sec:reduction}, and the background subtraction eradicates any significant diffuse structure away from the Galactic plane. 

\subsection{Catalogue combination and filtering}
\label{sec:catalog_filter}
The source tables for each image were concatenated into one all-southern-sky source catalogue, covering declinations $ \delta\leq 0^\circ $ for all right ascensions. 
For some right ascensions, sources are found outside the nominal declination boundary. 
There are 118 such sources with $ \delta>0^\circ $ that are included in the final catalogue and are used throughout the catalogue verification.

\begin{figure*}[!htbp]
\centering
\includegraphics[width=0.95\textwidth]{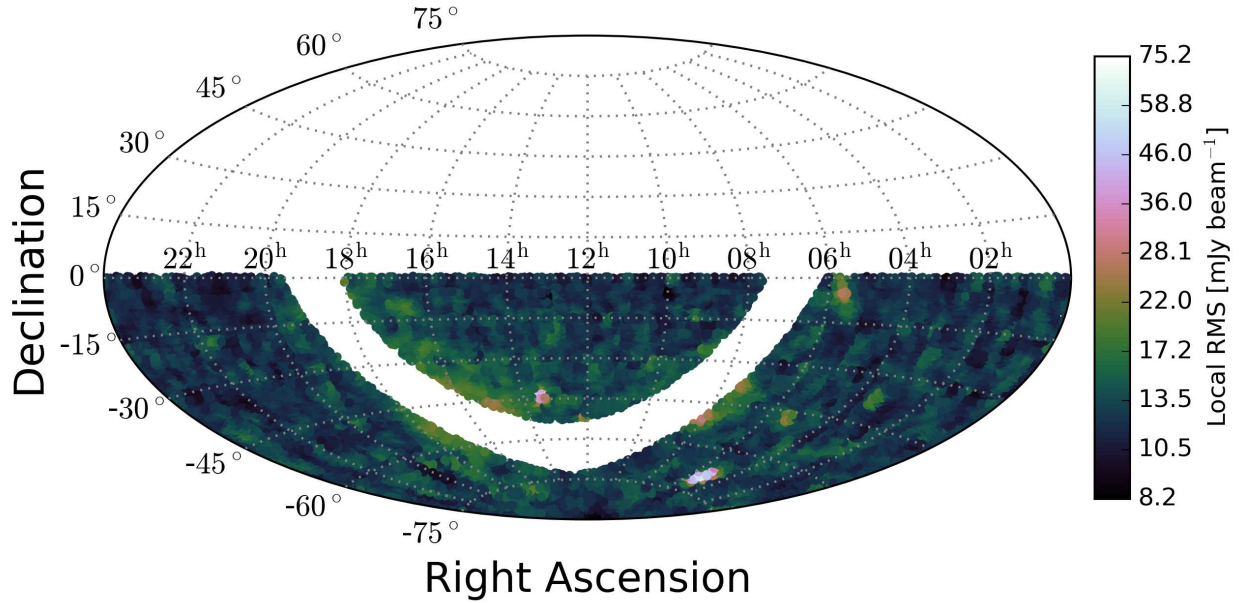}
\caption{An Aitoff projection RMS noise map of the sky area covered by \spass, including the $ |b|<10^\circ $ cut.
The mean local RMS noise for sources in the catalogue is $ \approx \rms\mathrm{\,mJy\,beam^{-1}} $ with notable exceptions being Centaurus A and the Large Magellanic Cloud which have local RMS values $ \sim 6 $ times the mean.}
\label{fig:sky_coverage}
\end{figure*}

The tile images at each declination strip overlap the next lowest declination strip by $ \sim 50\mathrm{\,arcmin} $. 
The overlap in right ascension varies with declination, ranging from $ \sim 1^\circ $ at the equator to $ \sim 10^\circ $ at $ \delta\approx -82^\circ $. 
Due to the overlap, the combined table contained multiple detections of several thousand sources. 
For each source that was detected multiple times, only the detection with the lowest RMS noise was retained in the final source catalogue. 

Sources near the Galactic plane ($ |b|<10^\circ $) were removed.
This conservative exclusion region was chosen because even though a median filter was applied to the scans, the Galactic plane would require a different analysis and catalogue creation pipeline due to the high source density and incomplete removal of large scale structures. 
This region will be examined in an upcoming paper. 

The final source catalogue contains \nsrc\ extragalactic sources and covers a sky area of approximately $\spasssky\mathrm{\,deg^2} $ (see Figure~\ref{fig:sky_coverage} for the RMS noise map).
Notable exceptions to the otherwise uniform sky noise level are: Centaurus A, the Large Magellanic Cloud and areas near the Galactic plane. 

The Stokes \textit{I} source catalogue format is outlined in Section~\ref{sec:cat_format} and an example selection of sources can be found in Tables~\ref{tab:spass_cat1}~and~\ref{tab:spass_cat2}.

\subsubsection{Resolved \spass\ sources}
\label{sec:resolved_srcs}
Given the \spass\ beam size, we would expect that few sources outside the Galactic plane will be partially or fully resolved with angular size $ \gtrsim 11\mathrm{\,arcmin} $. 

The ability to determine whether a source is resolved typically depends on the signal-to-noise of the source, where low signal-to-noise sources are much more difficult to constrain with an elliptical Gaussian. 
Using the fitted major and minor axes ($ a $ and $ b $) to estimate the source extent and we can determine whether the source is truly resolved. 
This assumes that all sources are well fit, thus any spurious fitting errors will produce nonsensical results. 

To assess how many sources are resolved, we define the extent of a source as
\begin{equation}
\zeta = \frac{ab}{a_\mathrm{psf}b_\mathrm{psf}},
\label{eq:extent}
\end{equation}
where $ a_\mathrm{psf} $ and $ b_\mathrm{psf} $ are the major and minor axis for the local point spread function.
In the case of \spass\, $ a_\mathrm{psf}\equiv b_\mathrm{psf}=645\mathrm{\,arcsec} $. 
The error in the extent is calculated by summing the fractional errors in $ a $ and $ b $ in quadrature, i.e. $ (\Delta\zeta/\zeta)^2\approx(\Delta a/a)^2+(\Delta b/b)^2 $.

\begin{figure}[!htbp]
\centering
\includegraphics[width=0.48\textwidth]{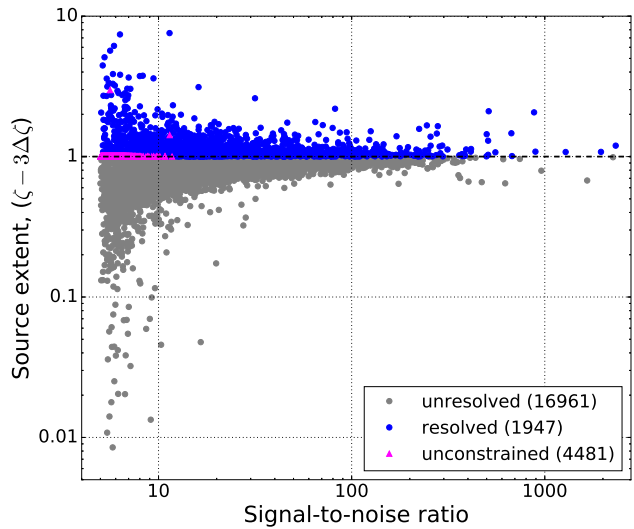}
\caption{\spass\ source extents ($ \zeta-3\Delta\zeta $) as a function of signal-to-noise ratio. 
Magenta triangles represent sources with unconstrained source size errors (i.e. $ \Delta a =-1 \mathrm{\,arcsec}$ or $ \Delta b=-1\mathrm{\,arcsec} $ in Table~\ref{tab:spass_cat2}).
Resolved sources are depicted as blue circles and unresolved sources are shown as grey circles.
The catalogue consists of $ 8\% $ resolved sources and $ 73\% $ unresolved sources, with the remaining $ 19\% $ having unconstrained source size errors. }
\label{fig:source_extents}
\end{figure}

A source is resolved at the $ 3\sigma $ level ($ \approx 99.7\% $ confidence assuming Gaussian statistics) if $ \left(\zeta-3\Delta\zeta\right)\geq 1 $, otherwise the source is unresolved. 
Figure~\ref{fig:source_extents} identifies three source categories: resolved, unresolved and unconstrained. 
Unconstrained sources are those for which \textsc{aegean} has been unable to determine errors in the semi-major ($ a $) or semi-minor ($ b $) axes.
Resolved and unresolved \spass\ source numbers are calculated from the total source catalogue minus those sources with unconstrained source size errors. 

Resolved sources comprise $ \sim 8\%$ of the total number of catalogued sources, while unresolved and unconstrained sources contribute $ \sim 73\% $ and $ \sim 19\% $. 
We consider all \nsrc\ sources, regardless of whether they are resolved or not, for the verification analysis.

\section{VERIFICATION}
\label{sec:analysis}
The \spass\ source catalogue consists primarily of compact sources. 
In order to assess the quality of the final catalogue, a number of tests have been performed. 

We analyse the internal catalogue flux density distribution and the average source spectra with respect to PMN counterparts at $ 4.8\mathrm{\,GHz} $ and PKSCAT90 counterparts at $ 2.7\mathrm{\,GHz} $. 
The catalogue astrometry, completeness and reliability are also examined in this section.

\subsection{Flux density scale}
\label{sec:flux}
The $ 16\mathrm{\,cm} $ ($ 2.1\mathrm{\,GHz} $) ATCA calibrator catalogue has high resolution ($ \sim 6\mathrm{\,arcsec} $, assuming $ 6\mathrm{\,km} $ array configuration), with sources selected to be compact and (mostly) have no other nearby source within \mbox{$ \sim 11\mathrm{\,arcmin} $}. 
ATCA calibrators were chosen for comparison with \spass\ sources because they provide an independent and accurate measurement of source flux densities.

The \spass\ catalogue was cross-matched with a list of calibrators\footnote{The compiled list of sources was created from accessing http://www.narrabri.atnf.csiro.au/calibrators/ on 02/02/2016.} for ATCA. 
The flux density limit for both source lists was restricted to $ S_\mathrm{peak}>500\mathrm{\,mJy\,beam^{-1}} $. 
The two catalogues were cross-matched symmetrically based on sky position with a $ 300\mathrm{\,arcsec} $ cross-matching radius, taking only the best matches. 
The cross-matching included all \spass\ sources, including those outside the nominal $ \delta\leq 0^\circ $ boundary. 
This produced a matched list containing 363 sources. 

After scaling the ATCA flux density to $ 2.3\mathrm{\,GHz} $ assuming a spectral index of $ -0.7 $, we calculate the ratio of the \spass\ to ATCA source flux density. 
The median flux density ratio is $ 1.04 \pm 0.01 $, which is consistent with unity given the \spass\ absolute flux calibration uncertainty (see Section~\ref{sec:observations}) and that errors in the ATCA flux measurements are not included.

\begin{figure}
\centering
\includegraphics[width=0.49\textwidth]{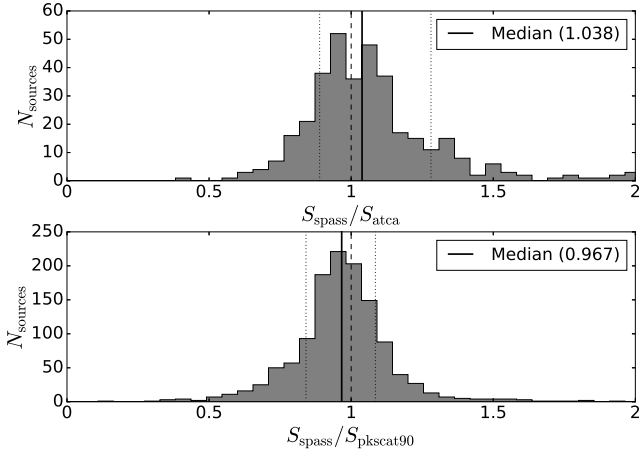}
\caption{Flux density ratio distributions for sources brighter than $ S_\mathrm{peak}>500\mathrm{\,mJy\,beam^{-1}} $. 
\textit{Top}: ATCA ($ 2.1\mathrm{\,GHz} $, extrapolated to $ 2.3\mathrm{\,GHz} $) and \spass\ source flux ratios. 
The median ratio, with standard error is $ 1.04\pm0.01 $.
\textit{Bottom}: PKSCAT90 ($ 2.7\mathrm{\,GHz} $, extrapolated to $ 2.3\mathrm{\,GHz} $) and \spass\ source flux ratios. 
The median ratio, with standard error is $ 0.967\pm0.003 $.
The median is displayed as a solid black line, a ratio of unity is marked by the dashed line, and the dotted lines represent the $ 1\sigma $ confidence levels.}
\label{fig:flux_ratio}
\end{figure}

The same analysis was also conducted using the PKSCAT90 $ 2.7\mathrm{\,GHz} $ fluxes.
This comparison has the benefit that both surveys were produced with the same instrument at similar frequencies and therefore with comparable resolution elements, reducing cross-matching confusion. 
The cross-matched list contains 1,232 sources. Scaling the PKSCAT90 fluxes to $ 2.3\mathrm{\,GHz} $, the median ratio of \spass\ to PKSCAT90 flux densities is $ 0.967 \pm 0.003 $.
This is again consistent within the \spass\ absolute flux calibration uncertainty of $ 10\% $. 

The distribution of ratios is expected to centre around unity. The results from cross-matching to both reference catalogues are plotted in Figure~\ref{fig:flux_ratio}. 
The dashed line indicates a ratio of 1, the solid line is the measured median ratio value and the dotted lines are the $ 1\sigma $ confidence levels. 
Given that both distributions have peaks consistent with unity, we assert that the \spass\ flux density scale is reliable to within the $ 10\% $ uncertainty.

\subsection{Spectral index distribution}
\label{sec:spec_ind}
To further test the accuracy of the \spass\ flux density scale, we examine the spectral index distribution between \spass\ at $ 2.3\mathrm{\,GHz} $ and PMN at $ 4.8\mathrm{\,GHz} $. 
PMN was chosen as its resolution ($ 5\mathrm{\,arcmin} $) is comparable to that of \spass\ ($ \sim 11\mathrm{\,arcmin} $), reducing cross-matching issues.
Spectral indices for 772 sources were calculated by cross-matching the \spass\ and PMN catalogues with a search radius of $ 300\mathrm{\,arcsec} $ and selecting only \spass\ sources brighter than $ 500\mathrm{\,mJy\,beam^{-1}} $. 

Caution should be taken when interpreting any individual source spectral index information for \spass\ and PMN. 
The surveys are separated by decades and source time-variability may result in drastic changes in observed spectral index properties, misrepresenting the true source spectral index.

\begin{figure}
\includegraphics[width=0.48\textwidth]{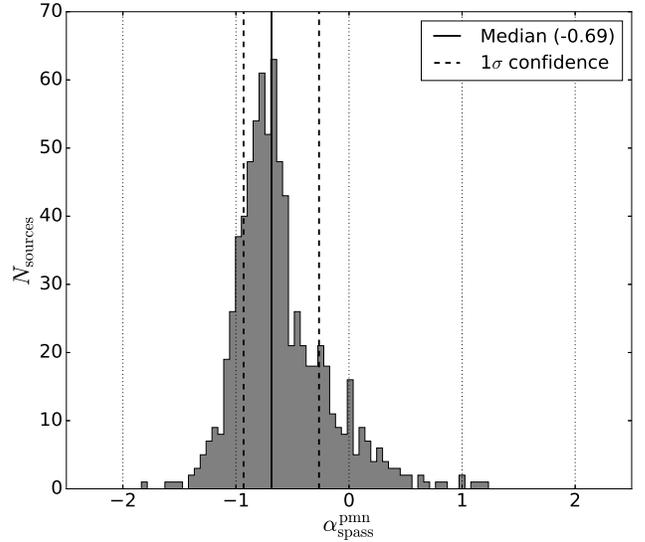}
\caption{Spectral index distribution between $ 2.3\mathrm{\,GHz} $ and $ 4.8\mathrm{\,GHz} $ for all \spass\ sources with $ S_\mathrm{peak}>500\mathrm{\,mJy\,beam^{-1}} $ and a counterpart in PMN. 
The solid black line identifies the median spectral index ($ \alpha_\mathrm{spass}^\mathrm{pmn}=-0.69\pm 0.02 $) and the black dashed lines represent the $ 1\sigma $ confidence interval ($ -0.93 $ to $ -0.26 $). 
Note the extended tail of flat and inverted spectral indices.}
\label{fig:alpha_hist}
\end{figure}

Figure~\ref{fig:alpha_hist} shows the distribution of spectral indices, with a median value (solid black line) of $\alpha_\mathrm{med}=-0.69\pm 0.02 $. 
The $ 1\sigma $ confidence interval (dashed black lines) spans spectral index values of $ -0.93 $ to $ -0.26 $. 
There are very few source populations that can achieve a spectral index of $ \alpha<-2 $, however a spectral index of $ -0.5\lesssim\alpha $ is not uncommon (e.g. blazars and QSOs). 
The tail of sources with spectral indices $ \alpha\gtrsim -0.5$, visible in Figure~\ref{fig:alpha_hist}, is therefore not unexpected. 
A similar distribution is observed independently by \citet{2016ApJLamee} using only a sample of $ \sim 500 $ \spass\ Stokes \textit{I} sources and cross-matching with NVSS.

The extended tail could be evidence for two source populations being partially resolved. 
In comparison to the Australia Telescope 20GHz Survey (AT20G; \citealp{2010MNRASMurphy}) spectral index distribution, where there is no clean distinction between source populations, it seems more likely that \spass\ is observing a single population with an extended ``flat'' spectrum tail. 

\subsection{Astrometry}
\label{sec:astrometry} 
The median signal-to-noise ratio for an \spass\ source is $ \mathrm{SNR}\sim 10 $. 
For sources with $ \mathrm{SNR}\sim 10 $, the mean position error, which accounts for the background noise, is $ \Delta\theta\sim 35\mathrm{\,arcsec} $ (using equations 20 and 21 from \citealp{1997PASPCondon}). 
The Gaussian fitting errors in RA and Dec (columns 4 and 5) that \textsc{aegean} calculates are consistent with the description given by \citet{1997PASPCondon}, assuming a synthesised beam of $ 645\mathrm{\,arcsec} $ and a pixel spacing of $ 0.05^\circ $.
We expect the mean errors in the right ascension (RA) and declination (Dec) for the entire catalogue to be approximately this value.

We cross-matched the \spass\ catalogue with the SUMSS and NVSS catalogues, chosen for their excellent astrometry. 
Bright source ($ S_\nu \gtrsim 5\mathrm{\,mJy\,beam^{-1}}  $) positions in NVSS are accurate to within $ (\epsilon_\alpha,\epsilon_\delta) = (0.45,0.56) \mathrm{\,arcsec}$ \citep{1998AJCondon}. 
The SUMSS catalogued sources have mean offsets from their cross-match with NVSS of $ \langle\Delta\alpha\rangle=-0.59\pm 0.07\mathrm{\,arcsec} $ and $ \langle\Delta\delta\rangle=-0.30\pm 0.08\mathrm{\,arcsec} $ \citep{2003MNRASMauch}.
Using a cross-matching radius of $ 10.75\mathrm{\,arcmin} $, we retrieve the average astrometric offsets for \spass\ sources.

\begin{figure}
\centering
\includegraphics[width=0.48\textwidth]{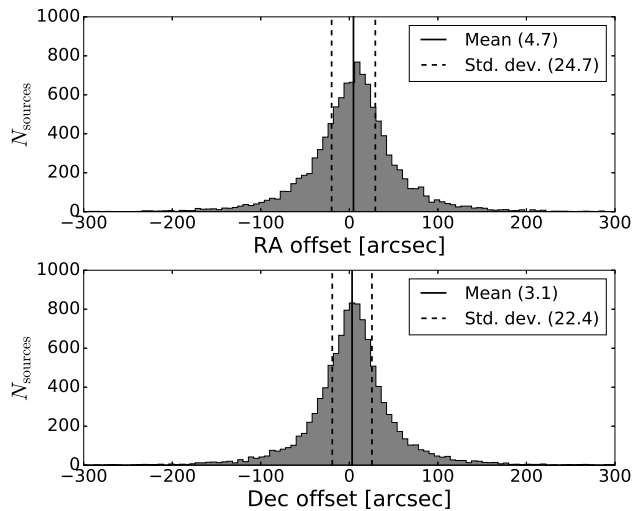}
\caption{Astrometric offset distributions from cross-matching \spass\ with SUMSS. 
The mean offset in RA is $ 4.7\mathrm{\,arcsec} $ and $ 3.1\mathrm{\,arcsec} $ in Dec.
The solid black line represents the distribution mean and the dashed lines identify the standard deviation.}
\label{fig:sumss_astrometry}
\end{figure}

\begin{figure}
\centering
\includegraphics[width=0.48\textwidth]{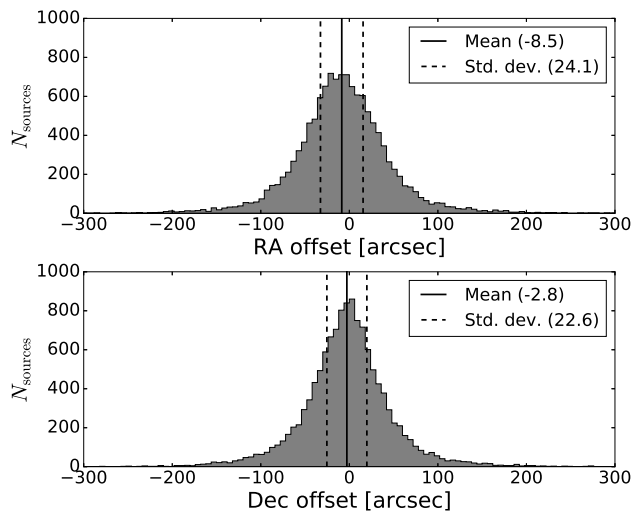}
\caption{Astrometric offset distributions from cross-matching \spass\ with NVSS. 
The mean offset in RA is $ -8.5\mathrm{\,arcsec}$ and $ -2.8\mathrm{\,arcsec} $ in Dec.
The solid black line represents the distribution mean and the dashed lines identify the standard deviation.}
\label{fig:nvss_astrometry}
\end{figure}

Cross-matching with SUMSS we find that the offsets are $ \langle\Delta\alpha\rangle=4.7\pm 24.7\mathrm{\,arcsec} $ and $ \langle\Delta\delta\rangle=3.1\pm 22.4\mathrm{\,arcsec} $ (see Figure~\ref{fig:sumss_astrometry}). 
Cross-matching with NVSS we find that the offsets are $ \langle\Delta\alpha\rangle=-8.5\pm 24.1\mathrm{\,arcsec} $ and $ \langle\Delta\delta\rangle=-2.8\pm 22.6\mathrm{\,arcsec} $ (see Figure~\ref{fig:nvss_astrometry}).


Cross-matching between catalogues with vastly different angular resolutions and much higher source densities, false matches can become an issue. 
Given the source densities of SUMSS and NVSS ($ \sim 21\mathrm{\,deg^{-2}} $, $ \sim 52\mathrm{\,deg^{-2}} $), and the size of the \spass\ beam ($ \mathrm{FWHM}=10.75\mathrm{\,arcmin} $), we could expect $ \sim 0.5 $ SUMSS sources, and $ \sim 1.3 $ NVSS sources per \spass\ beam. 
Consequently, this could lead to spurious cross-matches between unrelated sources and would increase the spread of astrometric offsets. 

Overall the astrometry for \spass\ sources has no net systematic offset and the errors are in agreement with the estimated $ \Delta\theta\sim 35\mathrm{\,arcsec} $. 
Note that individual source position errors are a function of signal-to-noise.

\subsection{Completeness}
\label{sec:completeness}
To calculate the catalogue completeness only \spass\ images that do not contain the Galactic plane were selected. 
The selection criterion was that the image centre Galactic latitude be more than $ 15^\circ $ away from the Galactic plane (i.e. $ |b_\mathrm{centre}|\geq 15^\circ $). This resulted in 80 of the original 107 images being used for this analysis.

To estimate the completeness, 200 simulated sources were injected, over a flux density range of $ 0.01\text{--}10\mathrm{\,Jy} $, into each of the selected \spass\ images.
The background and noise maps from the original images (i.e. before simulated source injection) were used with the simulated maps and processed by \textsc{aegean} in the same manner as when creating the source catalogue (see Section~\ref{sec:construction}).
This ensured that the background and noise properties of each simulated image were identical to those of the original maps.

The completeness ($ C_i $) for each image, $ i $, was calculated by counting the number of simulated sources detected ($ D_i $) versus the number injected into each image at each flux density bin (i.e. $ C_i(S_\nu)=D_i(S_\nu)/200 $). 
These completeness values were then combined to calculate the completeness for the entire catalogue.

\begin{figure}
\centering
\includegraphics[width=0.48\textwidth]{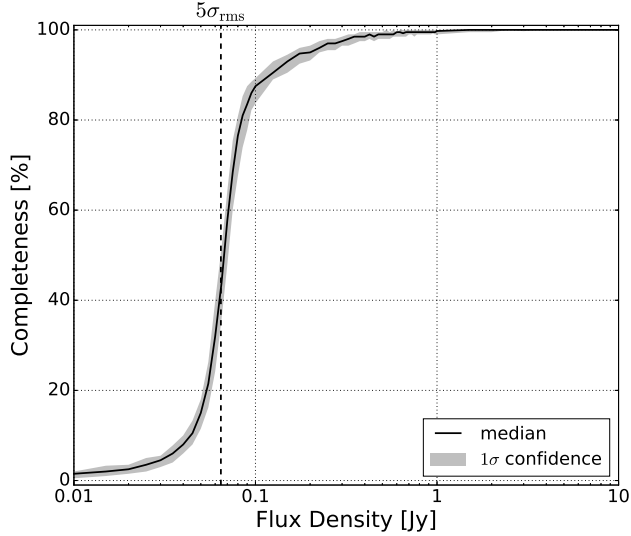}
\caption{The \spass\ catalogue median completeness (solid black line) and the $ 67\% $ confidence interval (shaded grey). The $ 5\sigma_\mathrm{rms} $ cut-off is indicated by the vertical dashed line.}
\label{fig:median_comp}
\end{figure}

In Figure~\ref{fig:median_comp}, the median completeness has been plotted for the catalogue, with the shaded region representing the $ 1\sigma $ confidence interval. 
The catalogue SNR cut-off, $ 5\sigma_\mathrm{rms} $ is plotted as a dashed black line for reference. 
The catalogue achieves a completeness of $>95\% $ at $ 0.225\mathrm{\,Jy} $ and is more than $ 99\% $ complete at $ 0.5\mathrm{\,Jy} $. 
The catalogue is $ 100\% $ complete above flux densities of $ 1\mathrm{\,Jy} $ and far from the Galactic plane.

\subsection{Reliability estimate}
\label{sec:reliability}
The reliability was estimated by measuring the fraction of \spass\ sources that have a counterpart in SUMSS for declinations $ -89^\circ\leq\delta\leq-40^\circ $, and NVSS for declinations $ -40^\circ<\delta\leq -1^\circ $. 
As when examining the flux scale and spectral index distribution of \spass\ sources, we select only those sources with a peak flux above $ 500\mathrm{\,mJy\,beam^{-1}} $, which is approximately the \spass\ $ 99\% $ completeness limit.
If we assume that the sources are self-absorbed ($ S_\nu\propto\nu^{2.5} $, i.e. a worst case scenario), this corresponds to SUMSS and NVSS flux densities of $ 41\mathrm{\,mJy\,beam^{-1}} $ and $ 145\mathrm{\,mJy\,beam^{-1}} $ respectively -- well above the $ 99\% $ completeness limit for each survey.

In the \spass\ source catalogue, there are 550 sources in the SUMSS region and 1003 sources in the NVSS region with \spass\ flux densities above $ 500\mathrm{\,mJy\,beam^{-1}} $.
Cross-matching SUMSS to \spass\ with a matching radius of $ 300\mathrm{\,arcsec} $ we find that there are 517 sources above the defined flux density limit with an \spass\ source above $ 500\mathrm{\,mJy\,beam^{-1}} $.
Using the same cross-matching criteria, we find there are 945 NVSS sources above the defined flux density limit with an \spass\ source above $ 500\mathrm{\,mJy\,beam^{-1}} $.

The ratio of sources detected in the cross-match to the number of suitable sources in \spass\ gives an estimate for the reliability at $ 500\mathrm{\,mJy\,beam^{-1}} $. 
For SUMSS and NVSS, this corresponds to $ \sim 94\% $ reliability.

As a baseline, a mock catalogue was created from the \spass\ source catalogue by shifting each source RA and Dec by $ +0.5^\circ $.
Cross-matching this mock catalogue in the same way as above we find that there are 58 matches between SUMSS and \spass\, and 14 between NVSS and \spass.
This corresponds to a false matching rate between $ \sim 2\text{--}9\% $, implying that the actual source catalogue reliability could be as low as $ 85\% $ at $ 500\mathrm{\,mJy\,beam^{-1}} $.

We note the discussion in Section~\ref{sec:astrometry} about source density considerations when cross-matching S-PASS with SUMSS and NVSS. 
The resolution difference between \spass, SUMSS and NVSS makes it difficult to disentangle whether sources are truly matched to the appropriate counterpart. 
One \spass\ beam can contain many SUMSS or NVSS sources, which results in a misleading cross-match. 
Also, some sources within SUMSS and NVSS may not be matched correctly due to imaging artefacts, high local noise levels or complex source structure, so the catalogue reliability may well be higher than calculated here.
In order to provide a comprehensive reliability estimate, the issue of cross-matching with different resolution catalogues should be addressed.
This would require a sophisticated algorithm, taking into account more than just simple distance between sources, such as the Positional Update and Matching Algorithm (PUMA\footnote{https://github.com/JLBLine/PUMA}; \citealp{2017PASALine}).

\section{CATALOGUE FORMAT}
\label{sec:cat_format}
An example of the first 25 sources has been included in Tables~\ref{tab:spass_cat1} and ~\ref{tab:spass_cat2}. 
A description of each column in the catalogue is as follows.\\
\textit{Column} (1): the \spass\ source name, formatted as SPASS\_Jhhmmss$\pm$ddmmss.\\
\textit{Columns} (2) \& (3): the J2000 RA in \textit{hh:mm:ss} and the J2000 Dec in \textit{dd:mm:ss}.\\
\textit{Columns} (4) \& (5): the errors in RA and Dec in arcseconds as quoted by \textsc{aegean}.\\
\textit{Columns} (6) \& (7): the peak flux density and associated error in $ \mathrm{Jy\,beam^{-1}} $. Uncertainties do not include flux scaling errors.\\
\textit{Columns} (8) \& (9): the integrated flux density and associated error in $ \mathrm{Jy} $. Uncertainties do not include flux scaling errors.\\
\textit{Columns} (10) \& (11): the background level and local RMS value, as calculated by \textsc{bane}, in $ \mathrm{Jy\,beam^{-1}} $.\\
\textit{Columns} (12) \& (13): the major axis of the fitted elliptical Gaussian and associated error in arcseconds.\\
\textit{Columns} (14) \& (15): the minor axis of the fitted elliptical Gaussian and associated error in arcseconds.\\
\textit{Columns} (16) \& (17): the position angle of the fitted elliptical Gaussian (measured East from North) and associated error in degrees.\\
\textit{Columns} (18) \& (19): the residual mean and residual standard deviation from the fitting process in $ \mathrm{Jy\,beam^{-1}} $.

\section{SUMMARY}
\label{sec:summary}
Using \spass\ total intensity data, the first southern-sky extragalactic source catalogue at $ 2.3\mathrm{\,GHz} $ has been created, containing \nsrc\ radio sources. 

The \spass\ source catalogue covers $ \spasssky\mathrm{\,deg^2} $ of sky.
The internal flux scale is reliable to within the $ 10\% $ calibration uncertainty estimate.
The \spass\ source spectral index distribution is consistent with a population with a median spectral index of $ \alpha\approx -0.7 $ and a tail of flat and inverted spectrum sources.

Typical astrometric offsets are consistent with approximately $ 35\mathrm{\,arcsec} $, though individual source astrometric errors vary as a function of signal-to-noise.
The catalogue is $ 95\% $ complete at $ 225\mathrm{\,mJy} $ and is $ 100\% $ complete above $ 1\mathrm{\,Jy} $. 
Approximately $ 94\% $ of \spass\ sources with a peak flux density above $ 500\mathrm{\,mJy\,beam^{-1}} $ have a lower-frequency counterpart.
Given the difference in source densities between \spass\ and the compared catalogues, this number is difficult to correctly estimate and could be as low as $ 85\% $.

A variety of science applications are possible using the \spass\ catalogue, including source spectrum studies by cross-matching with similar all-sky surveys, such as the newly released GaLactic and Extragalactic All-sky MWA (GLEAM) survey catalogue (\citealp{2015PASAWayth}, \citealp{2017MNRASHurley-Walker}) or the Planck Catalogue of Compact Sources (PCCS; \citealp{2014A&APlanckCollab}). 
With a centre frequency in the range where Giga-hertz Peaked Spectrum sources \citep{1991ApJODea} are expected to exhibit a spectral turn over, \spass\ would be a valuable addition to wide-band studies of these objects (e.g. \citealp{2015ApJCallingham}). 

\begin{acknowledgements}
This work has been carried out in the framework of the S-band Polarisation All Sky Survey (\spass) collaboration. 
The Parkes radio telescope is part of the Australia Telescope National Facility, which is funded by the Commonwealth of Australia for operation as a National Facility managed by CSIRO. 
Parts of this research were conducted by the Australian Research Council Centre of Excellence for All-sky Astrophysics (CAASTRO), through project number CE110001020. 
This research has made use of the VizieR catalogue access tool, CDS, Strasbourg, France.
The original description of the VizieR service was published in A\&AS 143, 23. 
The Dunlap Institute is funded through an endowment established by the David Dunlap family and the University of Toronto. 
B.M.G. acknowledges the support of the  Natural Sciences and Engineering Research Council of Canada (NSERC) through grant RGPIN-2015-05948, and of the Canada Research Chairs program.
\end{acknowledgements}

\bibliographystyle{pasa-mnras}
\bibliography{references}

\renewcommand{\arraystretch}{1.1}
\begin{landscape}
\begin{table}
\centering
\caption{The first 25 sources from the \spass\ catalogue, ordered by increasing Dec (column 3). The columns are defined in Section~\ref{sec:cat_format}. Continued in Table~\ref{tab:spass_cat2}.}
\label{tab:spass_cat1}
\begin{tabular}{@{}ccccccccccc@{}}
\hline\hline
  \multicolumn{1}{c}{(1)} &
  \multicolumn{1}{c}{(2)} &
  \multicolumn{1}{c}{(3)} &
  \multicolumn{1}{c}{(4)} &
  \multicolumn{1}{c}{(5)} &
  \multicolumn{1}{c}{(6)} &
  \multicolumn{1}{c}{(7)} &
  \multicolumn{1}{c}{(8)} &
  \multicolumn{1}{c}{(9)} &
  \multicolumn{1}{c}{(10)} &
  \multicolumn{1}{c}{(11)} \\
  \multicolumn{1}{c}{\spass\ name} &
  \multicolumn{1}{c}{RA (J2000)} &
  \multicolumn{1}{c}{Dec (J2000)} &
  \multicolumn{1}{c}{$ \Delta\mathrm{RA} $} &
  \multicolumn{1}{c}{$ \Delta\mathrm{Dec} $} &
  \multicolumn{1}{c}{$ S_\mathrm{peak} $} &
  \multicolumn{1}{c}{$ \Delta S_\mathrm{peak}^\star $} &
  \multicolumn{1}{c}{$ S_\mathrm{int} $} &
  \multicolumn{1}{c}{$ \Delta S_\mathrm{int}^\star $} &
  \multicolumn{1}{c}{background} &
  \multicolumn{1}{c}{local rms} \\
  \multicolumn{1}{c}{} &
  \multicolumn{1}{c}{$ [\mathrm{h\ m\ s}] $} &
  \multicolumn{1}{c}{$ [^\circ\ '\ ''] $} &
  \multicolumn{1}{c}{$ [''] $} &
  \multicolumn{1}{c}{$ [''] $} &
  \multicolumn{1}{c}{$ [\mathrm{Jy\,beam^{-1}}] $} &
  \multicolumn{1}{c}{$ [\mathrm{Jy\,beam^{-1}}] $} &
  \multicolumn{1}{c}{$ [\mathrm{Jy}] $} &
  \multicolumn{1}{c}{$ [\mathrm{Jy}] $} &
  \multicolumn{1}{c}{$ [\mathrm{Jy\,beam^{-1}}] $} &
  \multicolumn{1}{c}{$ [\mathrm{Jy\,beam^{-1}}] $} \\
\hline
SPASS\_J051336-302741 & 05:13:36 & -30:27:41 & 3 & 10 & 1.46 & 0.02 & 1.80 & 0.03 & -0.0022 & 0.0149 \\
SPASS\_J051450-301711 & 05:14:50 & -30:17:11 & 3 & 10 & 0.11 & 0.02 & 0.12 & 0.02 & -0.0022 & 0.0148 \\
SPASS\_J052257-295758 & 05:22:57 & -29:57:58 & 12 & 12 & 0.29 & 0.01 & 0.26 & 0.01 & -0.0022 & 0.0126 \\
SPASS\_J060043-293520 & 06:00:43 & -29:35:20 & 55 & 54 & 0.10 & 0.01 & 0.10 & 0.01 & -0.0031 & 0.0165 \\
SPASS\_J052632-294358 & 05:26:32 & -29:43:58 & 58 & 59 & 0.07 & 0.01 & 0.07 & 0.01 & -0.0022 & 0.0122 \\
SPASS\_J054618-293123 & 05:46:18 & -29:31:23 & 17 & 17 & 0.26 & 0.01 & 0.28 & 0.02 & -0.0035 & 0.0145 \\
SPASS\_J054523-294020 & 05:45:23 & -29:40:20 & 17 & 17 & 0.09 & 0.01 & 0.06 & 0.01 & -0.0034 & 0.0145 \\
SPASS\_J050557-293104 & 05:05:57 & -29:31:04 & 13 & 16 & 0.29 & 0.02 & 0.26 & 0.02 & -0.0023 & 0.0151 \\
SPASS\_J052250-293313 & 05:22:50 & -29:33:13 & 47 & 53 & 0.09 & 0.01 & 0.10 & 0.02 & -0.0023 & 0.0126 \\
SPASS\_J051544-292649 & 05:15:44 & -29:26:49 & 67 & 69 & 0.07 & 0.01 & 0.07 & 0.01 & -0.0023 & 0.0132 \\
SPASS\_J053126-292511 & 05:31:26 & -29:25:11 & 55 & 41 & 0.09 & 0.01 & 0.10 & 0.02 & -0.0025 & 0.0129 \\
SPASS\_J054953-291618 & 05:49:53 & -29:16:18 & 39 & 30 & 0.15 & 0.02 & 0.16 & 0.02 & -0.0034 & 0.0157 \\
SPASS\_J053755-291853 & 05:37:55 & -29:18:53 & 34 & 17 & 0.17 & 0.01 & 0.24 & 0.02 & -0.0028 & 0.0133 \\
SPASS\_J052539-291724 & 05:25:39 & -29:17:24 & 55 & 56 & 0.07 & 0.01 & 0.07 & 0.01 & -0.0022 & 0.0122 \\
SPASS\_J050150-290913 & 05:01:50 & -29:09:13 & 27 & 42 & 0.12 & 0.02 & 0.10 & 0.02 & -0.0025 & 0.0149 \\
SPASS\_J051138-290700 & 05:11:38 & -29:07:00 & 32 & 55 & 0.11 & 0.01 & 0.15 & 0.02 & -0.0023 & 0.0129 \\
SPASS\_J050740-290837 & 05:07:40 & -29:08:37 & 61 & 67 & 0.08 & 0.01 & 0.08 & 0.01 & -0.0023 & 0.0138 \\
SPASS\_J050244-290357 & 05:02:44 & -29:03:57 & 59 & 65 & 0.09 & 0.01 & 0.09 & 0.01 & -0.0025 & 0.0148 \\
SPASS\_J050537-285603 & 05:05:37 & -28:56:03 & 5 & 7 & 0.69 & 0.01 & 0.73 & 0.02 & -0.0024 & 0.0141 \\
SPASS\_J052156-285618 & 05:21:56 & -28:56:18 & 9 & 10 & 0.37 & 0.01 & 0.37 & 0.01 & -0.0023 & 0.0127 \\
SPASS\_J052045-284853 & 05:20:45 & -28:48:53 & 9 & 10 & 0.19 & 0.01 & 0.16 & 0.01 & -0.0022 & 0.0126 \\
SPASS\_J054318-285226 & 05:43:18 & -28:52:26 & 17 & 16 & 0.24 & 0.01 & 0.22 & 0.02 & -0.0030 & 0.0141 \\
SPASS\_J051543-285400 & 05:15:43 & -28:54:00 & 21 & 26 & 0.15 & 0.01 & 0.12 & 0.01 & -0.0022 & 0.0129 \\
SPASS\_J053955-283959 & 05:39:55 & -28:39:59 & 3 & 3 & 1.20 & 0.01 & 1.08 & 0.01 & -0.0029 & 0.0134 \\
SPASS\_J050122-283450 & 05:01:22 & -28:34:50 & 56 & 64 & 0.08 & 0.01 & 0.08 & 0.01 & -0.0026 & 0.0145 \\
\hline\hline
\end{tabular}
\tabnote{$^a$We stress that the uncertainties in peak and integrated flux densities do not include any correction for flux scaling errors.}
\end{table}
\end{landscape}

\begin{landscape}
\begin{table}
\centering
\caption{Continuation of Table~\ref{tab:spass_cat1}. Columns 3 and 4 from Table~\ref{tab:spass_cat1} have been appended to provide a reference.}
\label{tab:spass_cat2}
\begin{tabular}{@{}cc|cccccccc@{}}
\hline\hline
  \multicolumn{1}{c}{(2)} &
  \multicolumn{1}{c|}{(3)} &
  \multicolumn{1}{c}{(12)} &
  \multicolumn{1}{c}{(13)} &
  \multicolumn{1}{c}{(14)} &
  \multicolumn{1}{c}{(15)} &
  \multicolumn{1}{c}{(16)} &
  \multicolumn{1}{c}{(17)} &
  \multicolumn{1}{c}{(18)} &
  \multicolumn{1}{c}{(19)} \\
  \multicolumn{1}{c}{RA} &
  \multicolumn{1}{c|}{Dec} &
  \multicolumn{1}{c}{$ a $} &
  \multicolumn{1}{c}{$ \Delta a^\dagger $} &
  \multicolumn{1}{c}{$ b $} &
  \multicolumn{1}{c}{$ \Delta b^\dagger $} &
  \multicolumn{1}{c}{$ \mathrm{PA} $} &
  \multicolumn{1}{c}{$ \Delta \mathrm{PA}^\dagger $} &
  \multicolumn{1}{c}{res. mean$ ^\ddagger $} &
  \multicolumn{1}{c}{res. std$ ^\ddagger $} \\
  \multicolumn{1}{c}{$ [\mathrm{h\ m\ s}] $} &
  \multicolumn{1}{c|}{$ [^\circ\ '\ ''] $} &
  \multicolumn{1}{c}{$ [\mathrm{arcsec}] $} &
  \multicolumn{1}{c}{$ [\mathrm{arcsec}] $} &
  \multicolumn{1}{c}{$ [\mathrm{arcsec}] $} &
  \multicolumn{1}{c}{$ [\mathrm{arcsec}] $} &
  \multicolumn{1}{c}{$ [^\circ] $} &
  \multicolumn{1}{c}{$ [^\circ] $} &
  \multicolumn{1}{c}{$ [\mathrm{Jy\,beam^{-1}}] $} &
  \multicolumn{1}{c}{$ [\mathrm{Jy\,beam^{-1}}] $} \\
\hline
05:13:36 & -30:27:41 & 761.2 & 8.4 & 674.3 & 3.0 & 8.5 & 0.1 & -0.0029 & 0.0272 \\
05:14:50 & -30:17:11 & 756.1 & 8.4 & 627.4 & 3.0 & -78.1 & 0.1 & -0.0029 & 0.0272 \\
05:22:57 & -29:57:58 & 624.6 & 10.9 & 614.2 & 10.5 & 0.5 & 2.0 & 0.0018 & 0.0111 \\
06:00:43 & -29:35:20 & 651.3 & -1.0 & 645.0 & -1.0 & 0.1 & -1.0 & 0.0001 & 0.0026 \\
05:26:32 & -29:43:58 & 649.7 & -1.0 & 645.0 & -1.0 & -90.0 & -1.0 & -0.0006 & 0.0026 \\
05:46:18 & -29:31:23 & 684.7 & 16.3 & 647.3 & 14.6 & 4.2 & 0.7 & -0.0007 & 0.0061 \\
05:45:23 & -29:40:20 & 558.8 & 15.9 & 521.3 & 14.2 & -15.3 & 0.7 & -0.0007 & 0.0061 \\
05:05:57 & -29:31:04 & 636.4 & 13.9 & 598.3 & 12.2 & 5.9 & 0.5 & -0.0015 & 0.0096 \\
05:22:50 & -29:33:13 & 756.5 & 82.5 & 664.1 & 50.8 & -15.8 & 1.1 & -0.0026 & 0.0029 \\
05:15:44 & -29:26:49 & 650.2 & -1.0 & 645.0 & -1.0 & 89.9 & -1.0 & 0.0003 & 0.0015 \\
05:31:26 & -29:25:11 & 774.0 & 48.6 & 614.7 & 64.7 & 87.4 & 0.6 & 0.0001 & 0.0021 \\
05:49:53 & -29:16:18 & 698.5 & 34.3 & 636.5 & 43.4 & -85.4 & 1.0 & -0.0003 & 0.0050 \\
05:37:55 & -29:18:53 & 980.7 & 16.0 & 607.9 & 35.6 & 88.3 & 0.2 & 0.0000 & 0.0070 \\
05:25:39 & -29:17:24 & 650.0 & -1.0 & 645.0 & -1.0 & 90.0 & -1.0 & -0.0002 & 0.0006 \\
05:01:50 & -29:09:13 & 647.3 & 49.7 & 536.4 & 24.4 & 15.1 & 0.4 & 0.0006 & 0.0027 \\
05:11:38 & -29:07:00 & 884.3 & 68.0 & 658.9 & 34.9 & -4.8 & 0.2 & -0.0011 & 0.0119 \\
05:07:40 & -29:08:37 & 650.9 & -1.0 & 645.0 & -1.0 & 89.9 & -1.0 & 0.0000 & 0.0004 \\
05:02:44 & -29:03:57 & 651.4 & -1.0 & 645.0 & -1.0 & 89.8 & -1.0 & -0.0000 & 0.0004 \\
05:05:37 & -28:56:03 & 718.6 & 6.2 & 613.1 & 4.8 & 5.5 & 0.1 & 0.0023 & 0.0166 \\
05:21:56 & -28:56:18 & 659.6 & 10.1 & 622.1 & 8.2 & 10.4 & 0.4 & -0.0019 & 0.0146 \\
05:20:45 & -28:48:53 & 594.3 & 10.1 & 576.1 & 8.3 & -6.6 & 0.4 & -0.0019 & 0.0146 \\
05:43:18 & -28:52:26 & 630.9 & 15.9 & 618.6 & 14.3 & -9.1 & 2.3 & -0.0014 & 0.0068 \\
05:15:43 & -28:54:00 & 630.4 & 39.9 & 546.5 & 16.1 & 24.0 & 0.5 & 0.0004 & 0.0038 \\
05:39:55 & -28:39:59 & 648.6 & 3.5 & 578.9 & 2.4 & 10.8 & 0.1 & -0.0241 & 0.0470 \\
05:01:22 & -28:34:50 & 651.2 & -1.0 & 645.0 & -1.0 & 0.0 & -1.0 & -0.0024 & 0.0024 \\
\hline\hline
\end{tabular}
\tabnote{$^\dagger$Errors of -1 indicate a fitting error or nearly-circular source, where major and minor axis errors and position angle are poorly defined.}
\tabnote{$^\ddagger$The residual mean and std. dev. from the fitting process (i.e. data subtract model). If a source has been well fitted, the residuals will be small.}
\end{table}
\end{landscape}

\end{document}